\documentclass[sn-mathphys-num]{sn-jnl}% Math and Physical Sciences Numbered Reference Style 
\usepackage{graphicx}%
\usepackage{multirow}%
\usepackage{amsmath,amssymb,amsfonts}%
\usepackage{mathtools}
\usepackage{amsthm}%
\usepackage{mathrsfs}%
\usepackage[title]{appendix}%
\usepackage{xcolor}%
\usepackage{textcomp}%
\usepackage{manyfoot}%
\usepackage{booktabs}%
\usepackage{algorithm}%
\usepackage{algorithmicx}%
\usepackage{algpseudocode}%
\usepackage{listings}%
\DeclareMathOperator{\Tr}{\mathrm{Tr}}
\newcommand{\dd}{\mathrm{d}}

\newcommand{\dv}[1]{\frac{\mathrm{d}}{\mathrm{d}#1}}

\usepackage{braket}

\usepackage[capitalize,noabbrev]{cleveref}
\crefname{equation}{Eq.}{Eqs.}
\Crefname{equation}{Equation}{Equations}
\crefname{section}{Sec.}{Secs.}
\Crefname{section}{Section}{Sections}

\newcommand{\e}{e}
\newcommand{\im}{i}

\newcommand{\HS}{\hat{H}_\mathrm{S}}
\newcommand{\HB}{\hat{H}_\mathrm{B}}
\newcommand{\HI}{\hat{H}_\mathrm{I}}
\newcommand{\HT}{\hat{H}_\mathrm{T}}
\newcommand{\Heff}{\hat{H}_\mathrm{eff}}
\newcommand{\Veff}{\hat{V}_\mathrm{eff}}

\newcommand{\rhoS}{\rho_\mathrm{S}}
\newcommand{\rhoB}{\rho_\mathrm{B}}
\newcommand{\rhoT}{\rho_\mathrm{T}}
\newcommand{\rhoI}{\rho_\mathrm{I}}
\newcommand{\rhoss}{\rho_\mathrm{ss}}

\newcommand{\TrS}{\Tr_\mathrm{S}}
\newcommand{\TrB}{\Tr_\mathrm{B}}

\newcommand{\tauS}{\tau_\mathrm{S}}
\newcommand{\tauB}{\tau_\mathrm{B}}
\newcommand{\tauR}{\tau_\mathrm{R}}
\newcommand{\tauH}{\tau_\mathrm{H}}

\theoremstyle{thmstyleone}%
\theoremstyle{thmstyletwo}%

\theoremstyle{thmstylethree}%

\begin{document}

\title[Strong Markov dissipation in driven-dissipative quantum systems]{Strong Markov dissipation in driven-dissipative quantum systems}


\author{\fnm{Takashi} \sur{Mori}}\email{mori@rk.phys.keio.ac.jp}

\affil{\orgdiv{Department of Physics}, \orgname{Keio University}, \orgaddress{\street{Hiyoshi}, \city{Yokohama}, \postcode{223-8522}, \country{Japan}}}


\abstract{The Lindblad equation, which describes Markovian quantum dynamics under dissipation, is usually derived under the weak system-bath coupling assumption. Strong system-bath coupling often leads to non-Markov evolution. The singular-coupling limit is known as an exception: it yields a Lindblad equation with an arbitrary strength of dissipation. However, the singular-coupling limit requires high-temperature limit of the bath, and hence the system ends up in a trivial infinite-temperature state, which is not desirable in the context of quantum control. In this work, it is shown that we can derive a Markovian Lindblad equation for an arbitrary strength of the system-bath coupling by considering a new scaling limit that is called \emph{the singular-driving limit}, which combines the singular-coupling limit and fast periodic driving. In contrast to the standard singular-coupling limit, an interplay between dissipation and periodic driving results in a nontrivial steady state.
}

\keywords{open quantum systems, quantum master equation, periodic driving}



\maketitle
\section{Introduction}

The control of quantum many-body systems far from equilibrium is a central topic in modern physics. For this purpose, well-designed periodic driving has been utilized in the \emph{Floquet engineering}, which has shown growing interest in recent years~\citep{Bukov2015_review, Eckardt2017_review, Oka2019_review}.
It has also been recognized that dissipation due to the coupling to a bath can also be used to create novel phases of matter, which triggers an active research on the \emph{dissipation engineering}~\citep{Diehl2008, Verstraete2009}.

The theory of open quantum systems provides us a framework to incorporate the effect of dissipation~\citep{Breuer_text}.
In the Markovian regime, the dynamics of a dissipative quantum system is simply described by a Lindblad equation (or a GKSL equation named after Gorini, Kossakowski, Sudarshan~\citep{Gorini1976}, and Lindblad~\citep{Lindblad1976}).
However, the Lindblad equation is usually valid only for weak system-bath couplings, which strongly limits the potential utility of the dissipation engineering.
In the strong-coupling regime, on the other hand, non-Markovian processes significantly affect the dynamics~\citep{DeVega2017_review}, and the time-evolution equation becomes much more complicated.
As a result, it is in general difficult to theoretically predict what happens in the strong-coupling regime, which is not a desirable feature in the context of quantum control.

In this paper, we show that the interplay of strong coupling to a fast bath and high-frequency periodic driving can generate strong yet Markovian dissipation.
In such a situation, a simple Lindblad equation with a nontrivial steady state is derived for strong system-bath couplings.
This result opens a new possibility of controlling a quantum many-body system via strong Markovian dissipation.

In the following, we explain the setup in \cref{sec:setup}.
In \cref{sec:Lindblad}, we present a derivation of a Lindblad equation under the Born-Markov approximation for many-body open quantum systems periodically driven by external fields.
In the standard derivation of the Lindblad equation, the secular approximation (also called the rotating-wave approximation) is used~\citep{Breuer_text}, but it is known that the secular approximation is problematic for many-body systems.
For example, the Born-Markov-secular Lindblad equation cannot describe a nonequilibrium steady state with a finite current in the bulk~\citep{Wichterich2007}.
To overcome this difficulty, several derivations of the Lindblad equation have been proposed~\citep{Wichterich2007, Schaller2008, Benatti2010, Kirsanskas2018, Nathan2020, Becker2021}.
We here present a new derivation of the Lindblad equation without using the secular approximation.
Although the derived equation is identical to the universal Lindblad equation obtained in Ref.~\citep{Nathan2020}, our derivation clarifies the importance of a coarse-graining procedure, which is not so obvious in the original derivation.
We will see that there are two extreme situations in which the Born-Markov approximation is justified, which are discussed in \cref{sec:weak,sec:strong}, respectively.
It turns out that, in both cases, the steady state is described by a thermal Gibbs state $\e^{-\beta_\mathrm{eff}\Heff}/\TrS \e^{-\beta_\mathrm{eff}\Heff}$, where $\Heff$ is the effective system Hamiltonian (a time-averaged Hamiltonian in an appropriate rotating frame) and $\beta_\mathrm{eff}$ is the inverse effective temperature, which is not necessarily identical to the inverse temperature of the bath.
In \cref{sec:main}, we give a main result showing that strong Markovian dissipation can be realized by combining fast periodic driving with strong coupling to a fast bath.
We introduce a new scaling named as the \emph{singular-driving limit} in the derivation.
Remarkably, the derived Lindblad equation can have a nontrivial steady state.
We summarize our work and discuss future prospects in \cref{sec:summary}.

\section{Preliminaries}

\subsection{Setup}
\label{sec:setup}
Suppose a periodically driven system S in contact with a bath B.
The Hamiltonian of the total system T is written as
\begin{equation}
\HT(t)=\HS(t)+\HB+\HI,
\end{equation}
where $\HS(t)$, $\HB$, and $\HI$ are the Hamiltonian of the system, that of the bath, and the interaction Hamiltonian, respectively.
We assume $\HS(t)=\HS(t+T)$, where $T$ denotes the period of the driving field, and the corresponding frequency $\omega$ is defined as $\omega=2\pi/T$.
By assuming that the total system is isolated from the external environment, the density matrix $\rhoT(t)$ of the total system at time $t$ obeys the Liouville-von Neumann equation:
\begin{align}
\dv{t}\rhoT(t)=-\im [\HT(t),\rhoT(t)].
\label{eq:Liouville}
\end{align}

Throughout this work, we consider fast periodic driving, where the frequency $\omega$ is much larger than any characteristic local energy scale of the system S.
Fast periodic driving can have nontrivial effects in the case of strong driving or resonant driving, each of which is explained in the following.

\begin{description}
\item[Strong driving:]
The system Hamiltonian is given by
\begin{align}
\HS(t)=\hat{H}_0+\hat{V}(t)+\omega f(\omega t)\hat{A},
\label{eq:strong}
\end{align}
where $f(\theta)\in\mathbb{R}$ is a periodic function satisfying $f(\theta)=f(\theta+2\pi)$.
It is assumed that $\omega$ is large compared with any other characteristic local energy in $\hat{H}_0+\hat{V}(t)$.
The last term $\omega f(\omega t)\hat{A}$ is a strong periodic driving field, whereas $\hat{V}(t)$ is an additional periodic driving field with $\hat{V}(t)=\hat{V}(t+T)$.

\item[Resonant driving:]
The system Hamiltonian is given by
\begin{align}
\HS(t)=\hat{H}_0+\hat{V}(t)+\omega\hat{N},
\label{eq:resonant}
\end{align}
where $\hat{V}(t)=\hat{V}(t+T)$, and $\hat{N}$ is an operator such that every eigenvalue of it is integer.
Again, we assume that $\omega$ is large compared with a characteristic energy of $\hat{H}_0+\hat{V}(t)$.
Periodic oscillations with the period $T=2\pi/\omega$ can be resonant with the term $\omega\hat{N}$, which brings about nontrivial effects for large $\omega$.
\end{description}

In either case, it is convenient to move to a rotating frame via a periodic unitary transformation $\hat{U}(t)=\hat{U}(t+T)$ to extract nontrivial effects of periodic driving.
We choose $\hat{U}(t)=\e^{\im F(\omega t)\hat{A}}$ in the strong driving, where $F(\theta)=\int_0^\theta\dd\theta'\,f(\theta')$, whereas $\hat{U}(t)=\e^{\im\omega\hat{N}t}$ in the resonant driving.
In either case, \cref{eq:Liouville} is written in the following form in the rotating frame:
\begin{align}
\dv{t}\rhoT^\mathrm{R}(t)=-\im[\HT^\mathrm{R}(t),\rhoT^\mathrm{R}(t)],
\end{align}
where $\rhoT^\mathrm{R}=U(t)\rhoT(t)U^\dagger(t)$ and
\begin{align}
\HT^\mathrm{R}(t)=\HS^\mathrm{R}(t)+\HB+\HI^\mathrm{R}(t)
\end{align}
with $\HS^\mathrm{R}(t)=U^\dagger(t)(\hat{H}_0+\hat{V}(t))U(t)$ and $\HI^\mathrm{R}(t)=U^\dagger(t)\HI U(t)$.
We define an effective Hamiltonian as the time average of the system Hamiltonian in the rotating frame:
\begin{align}
\Heff=\frac{1}{T}\int_0^T\dd t\,\HS^\mathrm{R}(t)=\frac{1}{T}\int_0^T\dd t\, U^\dagger(t)\left(\hat{H}_0+\hat{V}(t)\right)U(t)
\end{align}
and the effective driving field as $\Veff(t)=\HS^\mathrm{R}(t)-\Heff$.
It should be noted that the interaction Hamiltonian depends on time in the rotating frame, which can generate nontrivial dissipative effects~\citep{Mori2023_review}.

Later, we always consider the problem in an appropriate rotating frame, and hence we omit the superscript R for notational simplicity.
%, but we still use $\Heff$ and $\Veff(t)$ instead of $\hat{H}_0$ and $\hat{V}(t)$, respectively, to emphasize that our Hamiltonian is modified by strong or resonant driving in a nontrivial way.

\subsection{Derivation of a Lindblad equation}
\label{sec:Lindblad}

The state of the system S is characterized by the reduced density matrix $\rhoS(t)=\TrB\rhoT(t)$, where $\TrB$ denotes the partial trace over the bath degrees of freedom.
We assume that the bath is in thermal equilibrium.
The theory of open quantum systems gives a framework of deriving an approximate equation of motion of $\rhoS(t)$ from the exact Liouville-von Neumann equation.

Without loss of generality, the interaction Hamiltonian $\HI(t)$ is expressed as
\begin{align}
\HI(t)=\sum_{i=1}^n\hat{X}_i(t)\otimes\hat{Y}_i=\sum_{i=1}^n\sum_{m=-\infty}^\infty\e^{\im m\omega t}\hat{X}_{i,m}\otimes\hat{Y}_i,
\label{eq:HI_decomposition}
\end{align}
where $\hat{X}_i(t)=\sum_{m=-\infty}^\infty\hat{X}_{i,m}\e^{\im m\omega t}$ is an Hermitian operator acting to the system S and $\hat{Y}_i$ is an Hermitian operator acting to the bath B.
Because of the Hermiticity of $\hat{X}_i(t)$, we have $\hat{X}_{i,m}^\dagger=\hat{X}_{i,-m}$.
It should be noted that we can recover the case of undriven systems by putting $\hat{X}_{i,m}=0$ for all $m\neq 0$.

We define
\begin{align}
\hat{X}_{i,m}(t)=\e^{\im\Heff t}\hat{X}_{i,m}\e^{-\im\Heff t}, \quad \hat{Y}_i(t)=\e^{\im\HB t}\hat{Y}_i\e^{-\im\HB t}.
\end{align}
The bath correlation function is given by
\begin{align}
\Phi_{ij}(t)=\TrB[\hat{Y}_i(t)\hat{Y}_j\rhoB],
\end{align}
where $\rhoB=\e^{-\beta\HB}/\Tr_B[\e^{-\beta\HB}]$ is the equilibrium state of the bath at the inverse temperature $\beta$.
More generally, we define multi-time correlation functions as
\begin{align}
\Phi_{i_1,i_2,\dots,i_l}(t_1,t_2,\dots,t_l)=\TrB[Y_{i_1}(t_1)Y_{i_2}(t_2)\dots Y_{i_l}(t_l)\rhoB].
\end{align}
We assume the existence of the bath correlation time $\tauB$ such that $\Phi_{i_1,i_2,\dots,i_l}(t_1,t_2,\dots,t_l)\approx 0$ if there is $k\in\{1,2,\dots,l\}$ such that $|t_k-t_j|>\tauB$ for any $j\neq k$.

Let us introduce other two timescales $\tauR$ and $\tauH$.
Here, $\tauR$ corresponds to the relaxation time due to dissipation.
Roughly speaking, $\tauR$ is given by the inverse of the dissipation strength, which is proportional to the square of the system-bath coupling in the Born-Markov regime (see \cref{sec:Lindblad}).
On the other hand, $\tauH$ is the heating time due to the external field $\Veff(t)$.
For high-frequency driving, $\tauH$ is usually exponentially small in frequency~\citep{Mori2016_rigorous, Kuwahara2016, Abanin2017_effective, Abanin2017_rigorous}.
In deriving a Lindblad equation, we make the following key assumptions:
\begin{itemize}
\item $\tauB\ll\tauR$. In this case, the Born-Markov approximation is justified.\footnote{In the Born approximation, we truncate the expansion in terms of the interaction Hamiltonian at the second order. It turns out that the smallness parameter $\epsilon$ of this expansion is given by $\epsilon=\sqrt{\tauB/\tauR}$.}
\item $\tauR\ll\tauH$. In this case, we can ignore the effect of the effective driving field $\Veff(t)$ since heating is immediately dissipated into the environment.
\end{itemize}
Under these assumptions, the reduced density matrix obeys the following Floquet-Redfield equation~\citep{Mori2023_review}
\begin{align}
\dv{t}\rhoS(t)=-\im[\Heff,\rhoS(t)]-\sum_{i=1}^n\sum_{m=-\infty}^\infty[\hat{X}_{i,m},\hat{R}_{i,-m}\rhoS(t)-\rhoS(t)\hat{R}_{i,m}^\dagger],
\label{eq:Redfield}
\end{align}
where
\begin{align}
\hat{R}_{i,m}=\sum_{j=1}^n\int_0^\infty \dd s\,\hat{X}_{j,m}(-s)\e^{-\im m\omega s}\Phi_{ij}(s).
\end{align}

\Cref{eq:Redfield} is still not of the Lindblad form, but we can find a Lindblad equation approximating \cref{eq:Redfield} without any further drastic approximation like the secular approximation.
A key idea leading to the Lindblad equation is to introduce a coarse graining time $\Delta t$ satisfying $\tauB\ll\Delta t\ll\tauR$.\footnote{A similar coarse-graining procedure was considered in previous works~\citep{Schaller2008, Benatti2010}, but the derivation presented here differs from the previous ones. The resultant Lindblad equations are therefore also different.
Remarkably, the Lindblad equation derived in the present work [see \cref{eq:ULE}] is described by only $n$ jump operators, whereas the previous ones have exponentially many jump operators for large system sizes.}
We do not take care about fast oscillations with frequencies $\Omega$ with $\Omega\Delta t\gg 1$.

Let us decompose the operator $\hat{X}_{i,m}$ as
\begin{align}
\hat{X}_{i,m}=\sum_\Omega\hat{X}_{i,m}[\Omega], \quad \hat{X}_{i,m}[\Omega]=\sum_{a,b: E_a-E_b=\Omega}\ket{a}\bra{a}\hat{X}_{i,m}\ket{b}\bra{b},
\end{align}
where $\ket{a}$ is an energy eigenstate with an energy eigenvalue $E_a$ of $\Heff$, i.e., $\Heff\ket{a}=E_a\ket{a}$.
We then find $\hat{X}_{i,m}(t)=\sum_\Omega\hat{X}_{i,m}[\Omega]\e^{\im\Omega t}$.
As a result, in the interaction picture, \cref{eq:Redfield} is written as
\begin{align}
\dv{t}\rhoI(t)=-\sum_{i,j=1}^n\sum_{m=-\infty}^\infty\sum_{\Omega,\Omega'} &\e^{\im(\Omega'-\Omega)t}
\left\{\int_0^\infty\dd s\, \e^{-\im(\Omega'+m\omega)s}\Phi_{ij}(s)[\hat{X}_{i,m}[\Omega]^\dagger,\hat{X}_{j,m}[\Omega']\rhoI]\right.\nonumber \\
&\left. -\int_0^\infty\dd s\,\e^{\im(\Omega+m\omega)s}\Phi_{ji}(s)^*[\hat{X}_{j,m}[\Omega'],\rhoI\hat{X}_{i,m}[\Omega]^\dagger]\right\},
\label{eq:Redfield_Omega}
\end{align}
where $\rhoI(t)=\e^{\im\Heff t}\rhoS(t)\e^{-\im\Heff t}$.
Let us define $\hat{\gamma}(\varepsilon)$ and $\hat{\eta}(\varepsilon)$ as matrices whose matrix elements are given by
\begin{align}
\gamma_{ij}(\varepsilon)=\int_{-\infty}^\infty\dd t\,\Phi_{ij}(t)\e^{-\im\varepsilon t}
\quad \text{and}\quad
\eta_{ij}(\varepsilon)=-\frac{i}{2}\int_{-\infty}^\infty\dd t\,\mathrm{sgn}(t)\Phi_{ij}(t)\e^{-\im\varepsilon t},
\label{eq:gamma_eta_explicit}
\end{align}
respectively.
It should be noted that $\hat{\gamma}(\varepsilon)$ and $\hat{\eta}(\varepsilon)$ are Hermitian at any $\varepsilon$: $\gamma_{ij}(\varepsilon)=\gamma_{ji}(\varepsilon)^*$ and $\eta_{ij}(\varepsilon)=\eta_{ji}(\varepsilon)^*$.
It is also shown that $\hat{\gamma}(\varepsilon)$ is positive semidefinite (i.e. every eigenvalue is non-negative).
The Fourier-Laplace transform of $\Phi_{ij}(t)$ is then expressed as
\begin{align}
\int_0^\infty \dd t\, \e^{-\im\varepsilon t}\Phi_{ij}(t)=\frac{1}{2}\gamma_{ij}(\varepsilon)+i\eta_{ij}(\varepsilon).
\label{eq:gamma_eta}
\end{align}
By substituting \cref{eq:gamma_eta} into \cref{eq:Redfield_Omega}, we obtain
\begin{align}
\dv{t}\rhoI(t)=&\sum_{ij=1}^n\sum_{m=-\infty}^\infty\sum_{\Omega,\Omega'}\e^{\im(\Omega'-\Omega)t}
\bigg\{-\im\eta_{ij}(\Omega'+m\omega)\left[\hat{X}_{i,m}[\Omega]^\dagger\hat{X}_{j,m}[\Omega'],\rhoI\right]
\nonumber \\
&\qquad\qquad\left.+\gamma_{ij}(\Omega'+m\omega)\left(\hat{X}_{j,m}[\Omega']\rhoI\hat{X}_{i,m}[\Omega]^\dagger-\frac{1}{2}\{\hat{X}_{i,m}[\Omega]^\dagger\hat{X}_{j,m}[\Omega'],\rhoI\}\right)\right\}
\nonumber \\
&+\sum_{ij=1}^n\sum_{m=-\infty}^\infty\sum_{\Omega,\Omega'}\e^{\im(\Omega'-\Omega)t}\Delta_{ij,m}(\Omega,\Omega')\left[\hat{X}_{j,m}[\Omega'],\rhoI\hat{X}_{i,m}[\Omega]^\dagger\right],
\label{eq:Redfield_decomp}
\end{align}
where
\begin{align}
\Delta_{ij,m}(\Omega,\Omega')=\frac{1}{2}\left[\gamma_{ij}(\Omega+m\omega)-\gamma_{ij}(\Omega'+m\omega)\right]-\im\left[\eta_{ij}(\Omega+m\omega)-\eta_{ij}(\Omega'+m\omega)\right].
\end{align}

We now perform the coarse graining.
In the coarse-grained timescale $\Delta t$, the factor $\e^{\im(\Omega'-\Omega)t}$ is averaged out when $|\Omega-\Omega'|\gg\Delta t^{-1}$.
Thus, the contribution from the terms with $|\Omega-\Omega'|\gg\Delta t^{-1}$ is negligible.
On the other hand, when $|\Omega-\Omega'|\lesssim\Delta t^{-1}\ll\tauB^{-1}$, we have\footnote{\Cref{eq:cg} is derived as follows.
Since the bath correlation function $\Phi_{ij}(t)$ vanishes for $t>\tauB$, we find $\int_0^\infty \dd s\,\e^{-\im(\Omega+m\omega)s}\Phi_{ij}(s)\approx\int_0^{\tauB}\dd s\,\e^{-\im(\Omega+m\omega)s}\Phi_{ij}(s)$.
By using the condition $|\Omega-\Omega'|\tauB\ll 1$, we obtain $\int_0^{\tauB}\dd s\,\e^{-\im(\Omega+m\omega)s}\Phi_{ij}(s)\approx\int_0^{\tauB}\dd s\,\e^{-\im(\Omega'+m\omega)s}\e^{-\im(\Omega-\Omega')s}\Phi_{ij}(s)\approx\int_0^{\tauB}\dd s\,\e^{-\im(\Omega'+m\omega)s}\Phi_{ij}(s)$.
We can thus conclude \cref{eq:cg}.}
\begin{align}
\int_0^\infty\dd s\, \e^{-\im(\Omega+m\omega)s}\Phi_{ij}(s)\approx\int_0^\infty\dd s\, \e^{-\im(\Omega'+m\omega)s}\Phi_{ij}(s),
\label{eq:cg}
\end{align}
which implies
\begin{align}
\hat{\gamma}(\Omega+m\omega)\approx\hat{\gamma}(\Omega'+m\omega) \quad \text{and} \quad
\hat{\eta}(\Omega+m\omega)\approx\hat{\eta}(\Omega'+m\omega).
\end{align}
As a result of the above argument, we can perform the following replacements for any pair of $\Omega$ and $\Omega'$ in \cref{eq:Redfield_decomp}:
\begin{align}
\left\{
\begin{aligned}
&\hat{\gamma}(\Omega+m\omega),\hat{\gamma}(\Omega'+m\omega)\to\hat{\gamma}^{1/2}(\Omega+m\omega)\hat{\gamma}^{1/2}(\Omega'+m\omega), \\
&\hat{\eta}(\Omega+m\omega),\hat{\eta}(\Omega'+m\omega)\to\hat{\eta}\left(\frac{\Omega+\Omega'}{2}+m\omega\right).
\end{aligned}
\right.
\label{eq:cg_replace}
\end{align}
Here, $\hat{\gamma}^{1/2}(\varepsilon)$ is the square root of $\hat{\gamma}(\varepsilon)$ satisfying $\sum_{k=1}^n\left[\hat{\gamma}^{1/2}(\varepsilon)\right]_{ik}\left[\hat{\gamma}^{1/2}(\varepsilon)\right]_{kj}=\gamma_{ij}(\varepsilon)$.

By using \cref{eq:cg_replace}, \cref{eq:Redfield_decomp} is reduced to the following \emph{non-secular Lindblad equation}:
\begin{align}
\dv{t}\rhoS(t)=-\im[\Heff+\hat{H}_\mathrm{LS},\rhoS]+\sum_{k=1}^n\sum_{m=-\infty}^\infty\left(\hat{L}_{k,m}\rhoS\hat{L}_{k,m}^\dagger-\frac{1}{2}\{\hat{L}_{k,m}^\dagger\hat{L}_{k,m},\rhoS\}\right),
\label{eq:ULE}
\end{align}
where
\begin{align}
\hat{H}_\mathrm{LS}=\sum_{ij}\sum_m\sum_{\Omega,\Omega'}\eta_{ij}\left(\frac{\Omega+\Omega'}{2}+m\omega\right)\hat{X}_{i,m}[\Omega]^\dagger\hat{X}_{j,m}[\Omega']
\label{eq:LS}
\end{align}
is the Lamb-shift Hamiltonian and
\begin{align}
\hat{L}_{k,m}=\sum_{i=1}^n\sum_\Omega\left[\hat{\gamma}^{1/2}(\Omega+m\omega)\right]_{ki}\hat{X}_{i,m}[\Omega]
\label{eq:jump}
\end{align}
is the jump operator.

In this way, as long as $\tauB\ll\tauR\ll\tauH$ holds, which is the same condition required for deriving the Floquet-Redfield equation, we can obtain the non-secular Lindblad equation.
We here emphasize that the non-secular Lindblad equation differs from the familiar Born-Markov-secular Lindblad equation.
In general, the secular approximation requires the condition $\delta E^{-1}\ll\tauR$, where $\delta E$ is a typical energy-level spacing of $\Heff$. 
In a many-body system, $\delta E$ is exponentially small in the system size, and hence the condition of $\delta E^{-1}\ll\tauR$ is unrealistic in macroscopic systems~\citep{Mori2023_review}.
On the other hand, \cref{eq:ULE} is derived without the secular approximation.

We should remark that \cref{eq:ULE} is essentially identical to the \emph{universal Lindblad equation} derived by \citet{Nathan2020}.
In our derivation, it is made clear that the coarse-graining procedure plays a key role, which is not so obvious in the original derivation.

So far, the timescale $\tauS$ for the intrinsic evolution of the system S has not appeared in the discussion.
Here, $\tauS^{-1}$ corresponds to a characteristic local energy scale of $\Heff$.
In the following, we discuss two extreme situations in which the condition $\tauB\ll\tauR$ holds in \cref{sec:weak,sec:strong}.

\subsection{Weak-coupling regime}
\label{sec:weak}

First, we consider the case of $\tauB,\tauS\ll\tauR$, which corresponds to a weak system-bath coupling.\footnote{We distinguish the weak-coupling regime discussed here from the ultra-weak coupling regime $\delta E^{-1}\ll\tauR$, in which the secular approximation is justified.}
In this case, dissipation is so weak that the dissipative term $\mathcal{D}\rhoS$ can be regarded as a small perturbation.
If the steady-state solution $\rhoss$ of \cref{eq:ULE} is expanded as $\rhoss=\sum_{k=0}^\infty\rho_k$, where $\rho_k$ represents the $k$th order term with respect to the perturbation, it is shown that the leading-order term $\rho_0$ is diagonal in the energy basis, i.e., $\rho_0=\sum_aP_a\ket{a}\bra{a}$, where $P_a\geq 0$ and $\sum_aP_a=1$.
Here, $\{P_a\}$ are determined by solving the following equations:
\begin{align}
\sum_{b(\neq a)}\left(W_{ab}P_b-W_{ba}P_a\right)=0,
\label{eq:Pauli}
\end{align}
where the ``transition rate'' $W_{ab}$ is defined as
\begin{align}
W_{ab}=\sum_{ij}\sum_m\gamma_{ij}(E_a-E_b+m\omega)\braket{a|\hat{X}_{j,m}|b}\braket{b|\hat{X}_{i,m}^\dagger|a},
\end{align}
which is non-negative because the matrix $\gamma_{ij}(\varepsilon)$ is positive semidefinite.

For weak dissipation, $\rhoss\approx\rho_0=\sum_aP_a\ket{a}\bra{a}$.
If the system described by $\Heff$ is chaotic, it is expected that it obeys the eigenstate thermalization hypothesis (ETH)~\citep{DAlessio2016_review, Mori2018_review}, which states that any energy eigenstate $\ket{a}$ is locally equivalent to a thermal Gibbs state $\rhoS^\mathrm{eq}=\e^{-\beta_a\Heff}/\TrS[\e^{-\beta_a\Heff}]$ at the inverse temperature $\beta_a$ corresponding to the energy $E_a$, i.e., $\beta_a$ is determined by the condition of $\TrS[\Heff\rhoS^\mathrm{eq}]=E_a$.
Thus, if the fluctuation of the energy in the state $\rho_0$ is subextensive, which is expected to hold in any realistic state of a macroscopic system, $\rho_0$ is locally equivalent to a thermal Gibbs ensemble at a certain temperature~\citep{Shirai2020} even though the detailed-balance condition is violated in \cref{eq:Pauli} under periodic driving~\citep{Breuer2000, Kohn2001}.\footnote{The condition that the populations $\{P_a\}$ are truely given by the Gibbs distribution $P_a\propto e^{-\beta E_a}$, where $\beta$ is the inverse temperature of the bath, is discussed in Refs.~\citep{Shirai2015,Shirai2016}.}
It means that weak dissipation just controls the temperature of the equilibrium system: we cannot go beyond equilibrium phenomena.
Thus, dissipation engineering is strongly limited in the weak-coupling regime.

\subsection{Strong-coupling regime}
\label{sec:strong}

Next, we consider the case of $\tauB\ll\tauS,\tauR$.
The condition of $\tauB\ll\tauS$ implies that the dynamics of the bath is much faster than that of the system of interest.
Since $\tauS\sim\tauR$ (or even $\tauR\ll\tauS$), dissipation is comparable or even stronger than a characteristic energy of the system S.

It should be noted that the condition of $\tauB\ll\tauS$ requires a high temperature of the bath.
Indeed, for an equilibrium bath, $\gamma_{ij}(\varepsilon)$ obeys the Kubo-Martin-Schwinger (KMS) relation
\begin{align}
\gamma_{ij}(\varepsilon)=\gamma_{ji}(-\varepsilon)\e^{-\beta\varepsilon}.
\label{eq:KMS}
\end{align}
This implies that $\gamma_{ij}(\varepsilon)$ should change when $\varepsilon$ changes by an amount of $\Delta\varepsilon\sim\beta^{-1}$.
Since $\gamma_{ij}(\varepsilon)\simeq\gamma_{ij}(\varepsilon+\Delta\varepsilon)$ as long as $|\Delta\varepsilon|\lesssim\tauB^{-1}$, we conclude $\beta\lesssim\tauB$.
From this argument, we find that $\tauB\ll\tauS$ implies $\beta\ll\tauS$, which means that the bath must have an infinitely high temperature.
As a consequence, the system S will eventually reach a boring infinite-temperature state, $\rhoss\propto\hat{I}_\mathrm{S}$, where $\hat{I}_\mathrm{S}$ is the identity operator acting to the Hilbert space of the system S.

The above conclusion is confirmed by considering a delta-correlated bath with $\Phi_{ij}(t)=\gamma_{ij}\delta(t)$.
\Cref{eq:KMS} requires $\beta=0$ and $\gamma_{ij}=\gamma_{ji}\in\mathbb{R}$.
In this case, the dissipative part of \cref{eq:ULE}, which is denoted by $\mathcal{D}\rhoS$, reads
\begin{align}
\mathcal{D}\rhoS=\sum_{ij=1}^n\sum_{m=-\infty}^\infty\gamma_{ij}\left(\hat{X}_{j,m}\rhoS\hat{X}_{i,m}^\dagger-\frac{1}{2}\{\hat{X}_{i,m}^\dagger\hat{X}_{j,m},\rhoS\}\right).
\label{eq:D_singular_coupling}
\end{align}
By using $\hat{X}_{i,m}=\hat{X}_{i,-m}^\dagger$, we can see that $\mathcal{D}\rhoS$ vanishes when $\rhoS\propto\hat{I}_\mathrm{S}$, and hence, the Lindblad equation has an infinite-temperature steady state.

Mathematically, a delta-correlated bath is realized in the \emph{singular-coupling limit}~\citep{Palmer1977}: the Hamiltonian of the total system is written as
\begin{align}
\HT=\HS+\lambda^{-2}\HB+\lambda^{-1}\HI
\end{align}
and taking the limit of $\lambda\to 0$ and $\beta\to 0$ with $\beta/\lambda^2$ held fixed.

The singular-coupling limit was first studied by \citet{Hepp1973}.
\citet{Palmer1977} rigorously proved, without any approximation such as the Born-Markov approximation, that the dynamics of a finite quantum system that is linearly coupled to a free-fermion bath is described by a Lindblad equation in the singular-coupling limit.

It should be remarked that a fine-tuned Lindblad equation in the singular-coupling limit can have a nontrivial steady state besides the infinite-temperature state.
For example, if one can design jump operators $\{\hat{X}_{i,m}\}$ in \cref{eq:D_singular_coupling} so that $\mathcal{D}(\ket{\psi}\bra{\psi})=0$ for an eigenstate $\ket{\psi}$ of $\Heff$, the Lindblad equation has a pure steady state $\ket{\psi}\bra{\psi}$.
Such a steady state is called a dark state in quantum optics and quantum-information theory~\citep{Kraus2008}, which is useful in the reservoir engineering.

\section{Strong Markovian dissipation leading to a nontrivial steady state}
\label{sec:main}

As we have already seen in the previous section, strong coupling to a fast bath with $\tauB\to +0$ (i.e. the singular coupling) in general leads to the trivial infinite-temperature state unless the system-bath coupling is fine-tuned.
In this section, we show that strong Markovian dissipation that leads to a nontrivial steady state is realized without fine tuning by combining singular coupling to a bath and fast periodic driving in an appropriate way, which is mathematically described by \emph{the singular-driving limit}.

\subsection{Singular-driving limit}

In the following, we consider the situation of
\begin{align}
\omega^{-1}\sim\tauB\ll\tauS,\tauR.
\end{align}
To realize this situation, we introduce a scaling parameter $\lambda$ such that
\begin{align}
\HT(t)=\Heff+\frac{1}{\lambda^2}\HB+\frac{1}{\lambda}\HI(t), \quad \omega=\frac{\tilde{\omega}}{\lambda^2}, \quad \beta=\lambda^2\tilde{\beta},
\label{eq:singular}
\end{align}
where $\tilde{\omega}$ and $\tilde{\beta}$ are the scaled frequency and the scaled inverse temperature.
We will take the limit of $\lambda\to+0$ with $\tilde{\omega}$ and $\tilde{\beta}$ held fixed.

Without the driving field, this limit reduces to the singular-coupling limit discussed in \cref{sec:strong}.
In \cref{eq:singular}, the driving frequency is also scaled with $\lambda$, which gives rise to nontrivial effects.
We shall call the limit of \cref{eq:singular} the \emph{singular-driving limit}.

In the singular-driving limit, the bath is delta-correlated.
The bath correlation function $\Phi_{ij}(t)$ is given by
\begin{align}
\Phi_{ij}(t)=\frac{1}{\lambda^2}\TrB\left[\e^{\im\HB t/\lambda^2}\hat{Y}_i\e^{-\im\HB t/\lambda^2}\hat{Y}_j\frac{\e^{-\tilde{\beta}\HB}}{\TrB[\e^{-\tilde{\beta}\HB}]}\right]\eqqcolon\frac{1}{\lambda^2}\tilde{\Phi}_{ij}\left(\frac{t}{\lambda^2}\right),
\end{align}
where
\begin{equation}
\tilde{\Phi}_{ij}(t)=\TrB\left[\e^{\im\HB t}\hat{Y}_i\e^{-\im\HB t}\hat{Y}_j\frac{\e^{-\tilde{\beta}\HB}}{\TrB[\e^{-\tilde{\beta}\HB}]}\right]
\end{equation}
is the rescaled bath correlation function.
We assume that $\tilde{\Phi}_{ij}(t)$ is a smooth function of $t$ and satisfies $\int_0^\infty\dd t\,|\tilde{\Phi}_{ij}(t)|<+\infty$.

In the limit of $\lambda\to+0$, we have, for $\varepsilon$ independent of $\lambda$,
\begin{align}
\int_0^\infty\dd t\,\Phi_{ij}(t)\e^{-\im(\varepsilon+m\omega)t}&=\int_0^\infty\dd t\,\tilde{\Phi}_{ij}(t)\e^{-\im\varepsilon\lambda^2t}\e^{-\im m\tilde{\omega}t}
\nonumber \\
&\to\int_0^\infty\dd t\,\tilde{\Phi}_{ij}(t)\e^{-\im m\tilde{\omega}t}\eqqcolon\frac{1}{2}\tilde{\gamma}_{ij}(\tilde{\omega})+\im\tilde{\eta}_{ij}(m\tilde{\omega}).
\end{align}
We therefore find
\begin{align}
\lim_{\lambda\to+0}\gamma_{ij}(\varepsilon+m\omega)=\tilde{\gamma}_{ij}(m\tilde{\omega})
\quad\text{and}\quad
\lim_{\lambda\to+0}\eta_{ij}(\varepsilon+m\omega)=\tilde{\eta}_{ij}(m\tilde{\omega}).
\end{align}
By substituting them into \cref{eq:ULE}, we obtain
\begin{align}
\dv{t}\rhoS=&-\im\left[\Heff+\sum_{ij}\sum_m\tilde{\eta}_{ij}(m\tilde{\omega})\hat{X}_{i,m}^\dagger\hat{X}_{j,m},\rhoS\right]
\nonumber \\
&+\sum_{ij}\sum_m\tilde{\gamma}_{ij}(m\tilde{\omega})\left(\hat{X}_{j,m}\rhoS\hat{X}_{i,m}^\dagger-\frac{1}{2}\{\hat{X}_{i,m}^\dagger\hat{X}_{j,m},\rhoS\}\right).
\label{eq:Lindblad_singular}
\end{align}
The KMS relation yields
\begin{align}
\tilde{\gamma}_{ij}(m\tilde{\omega})=\tilde{\gamma}_{ji}(-m\tilde{\omega})\e^{-\tilde{\beta}m\tilde{\omega}},
\label{eq:KMS_singular}
\end{align}
which implies that $\tilde{\gamma}_{ij}(m\tilde{\omega})$ is not a constant as long as $\tilde{\beta}\tilde{\omega}$ is nonzero, and therefore can generate a nontrivial nonequilibrium steady state.
In the derivation of \cref{eq:Lindblad_singular}, the weak coupling between the system S and the bath B has not been assumed.
\Cref{eq:Lindblad_singular} is valid even for the strong coupling regime, $\tauS\sim\tauR$, where the steady state is not necessarily described by a thermal Gibbs state $\e^{-\beta_\mathrm{eff}\Heff}/\TrS[\e^{-\beta_\mathrm{eff}\Heff}]$.

\subsection{A pedagogical example}
\label{sec:example}

We discuss a simple example of the Lindblad equation in the singular-driving limit.
Let us consider a resonantly driven spin 1/2 in contact with a boson bath, where the Hamiltonian of the total system in the static frame is given by
\begin{equation}
\HT(t)=h^z\hat{\sigma}^z+\omega\hat{\sigma}^z+\xi\cos(\omega t)\hat{\sigma}^x+\frac{1}{\lambda^2}\sum_k\omega_k\hat{a}_k^\dagger\hat{a}_k+\frac{1}{\lambda}\hat{\sigma}^x\sum_kg_k(\hat{a}_k+\hat{a}_k^\dagger),
\label{eq:HT_example}
\end{equation}
where $\hat{\sigma}^{\alpha}$ ($\alpha=x,y,z$) is the Pauli matrix, $\hat{a}_k$ ($\hat{a}_k^\dagger$) is the annihilation (creation) operator of bosons of mode $k$ in the bath, $\omega_k$ is the energy of a boson of mode $k$, and $g_k\in\mathbb{R}$ is the coupling constant between the system (the spin 1/2) and the boson of mode $k$.
We assume that $\omega>0$.

In the rotating frame generated by the unitary operator $\hat{U}(t)=e^{i\omega\hat{\sigma}^zt}$, the Hamiltonian is given in the form $\Heff+\Veff(t)+(1/\lambda^2)\HB+(1/\lambda)\HI(t)$ with
\begin{equation}
\left\{
\begin{aligned}
&\Heff=h^z\hat{\sigma}^z+\frac{\xi}{2}\hat{\sigma}^x, & &\Veff(t)=\frac{\xi}{2}(\hat{\sigma}^+\e^{2\im\omega t}+\hat{\sigma}^-\e^{-2\im\omega t})\\
&\HB=\sum_l\omega_k\hat{a}_k^\dagger\hat{a}_k, & &\HI(t)=(\hat{\sigma}^+\e^{\im\omega t}+\hat{\sigma}^-\e^{-\im\omega t})\sum_kg_k(\hat{a}_k+\hat{a}_k^\dagger),
\end{aligned}
\right.
\end{equation}
where $\hat{\sigma}^\pm=(\hat{\sigma}^x\pm i\hat{\sigma}^y)/2$.
According to \cref{eq:HI_decomposition}, we express the interaction Hamiltonian in the form $\HI(t)=(\hat{X}_1\e^{\im\omega t}+\hat{X}_{-1}\e^{-\im\omega t})\otimes\hat{Y}$, where $\hat{X}_1=\hat{\sigma}^+$, $\hat{X}_{-1}=\hat{\sigma}^-$, and $\hat{Y}=\sum_k g_k(\hat{a}_k+\hat{a}_k^\dagger)$.

By using the notation $\eta_\pm\coloneqq\tilde{\eta}(\pm\tilde{\omega})$ and $\gamma_\pm\coloneqq\tilde{\gamma}(\pm\tilde{\omega})$, \cref{eq:Lindblad_singular} is written as
\begin{equation}
\begin{aligned}
\dv{t}\rhoS(t)=&-\im\left[\Heff-\frac{\eta_+-\eta_-}{2}\hat{\sigma}^z,\rhoS\right]\\
&+\gamma_+\left(\hat{\sigma}^+\rhoS\hat{\sigma}^--\frac{1}{2}\{\hat{\sigma}^-\hat{\sigma}^+,\rhoS\}\right)
+\gamma_-\left(\hat{\sigma}^-\rhoS\hat{\sigma}^+-\frac{1}{2}\{\hat{\sigma}^+\hat{\sigma}^-,\rhoS\}\right),
\end{aligned}
\label{eq:Lindblad_singular_example}
\end{equation}
which is the Lindblad equation in the singular-driving limit.
By using the bath spectral density $J(\omega)=\sum_kg_k^2\delta(\omega-\omega_k)$, the constants $\gamma_\pm$ and $\eta_\pm$ are explicitly given by
\begin{equation}
\gamma_\pm=\pm \frac{2\pi J(\tilde{\omega})}{\e^{\pm\tilde{\beta}\tilde{\omega}}-1}
\label{eq:gamma_pm}
\end{equation}
and
\begin{equation}
\eta_\pm=\mathrm{P}\int_{-\infty}^\infty\dd\omega\,\frac{1}{\omega\mp\tilde{\omega}}\frac{J(|\omega|)\mathrm{sgn}(\omega)}{e^{\tilde{\beta}\omega}-1},
\label{eq:eta_pm}
\end{equation}
where $\mathrm{P}\int$ denotes the Cauchy principal value integral.
One can easily confirm that $\gamma_\pm$ satisfies \cref{eq:KMS_singular}, i.e., $\gamma_+/\gamma_-=e^{-\tilde{\beta}\tilde{\omega}}$.

The dissipator in \cref{eq:Lindblad_singular_example} has the following interpretation: the spin flips from down to up state with rate $\gamma_+$ and from up to down state with rate $\gamma_-$.
When $\gamma_+\neq\gamma_-$, a nontrivial steady state is realized.

Let us compare \cref{eq:Lindblad_singular_example} with the Lindblad equation obtained in the standard singular-coupling limit, in the latter of which we fix the driving frequency $\omega$ and take the limit of $\lambda\to +0$ with $\tilde{\beta}=\beta/\lambda^2$ held fixed.
This limiting procedure is equivalent to put $\tilde{\omega}=0$ in the singular-driving limit, and hence we have $\gamma_+=\gamma_-\eqqcolon\gamma$ and $\eta_+=\eta_-$.
We therefore have
\begin{align}
\dv{t}\rhoS(t)&=-\im[\Heff,\rhoS]
+\gamma\left(\hat{\sigma}^+\rhoS\hat{\sigma}^--\frac{1}{2}\{\hat{\sigma}^-\hat{\sigma}^+,\rhoS\}\right)
+\gamma\left(\hat{\sigma}^-\rhoS\hat{\sigma}^+-\frac{1}{2}\{\hat{\sigma}^+\hat{\sigma}^-,\rhoS\}\right)
\nonumber \\
&=-\im[\Heff,\rhoS]+\frac{\gamma}{2}\left(\hat{\sigma}^x\rhoS\hat{\sigma}^x-\rhoS\right)+\frac{\gamma}{2}\left(\hat{\sigma}^y\rhoS\hat{\sigma}^y-\rhoS\right).
\end{align}
In this equation, the transition from down to up state and its inverse transition occur with the same rate $\gamma$, and the steady state is the trivial infinite temperature state.
Only when the driving frequency scales as $\lambda^{-2}$, two transition rates $\gamma_\pm$ can have different values.

In the singular-driving limit, the dissipator does not depend on the system Hamiltonian $\Heff$, and hence \cref{eq:Lindblad_singular_example} is straightforwardly extended to many-body systems.
Instead of \cref{eq:HT_example}, we consider the following Hamiltonian of the total system in the periodic boundary condition:
\begin{equation}
\begin{aligned}
\HT(t)=&\sum_{i=1}^L\left[J\hat{\sigma}_i^z\hat{\sigma}_{i+1}^z+h^z\hat{\sigma}_i^z+\xi\cos(\omega t)\hat{\sigma}_i^x\right] \\
&+\frac{1}{\lambda^2}\sum_{i=1}^L\sum_k\omega_k\hat{a}_{i,k}^\dagger\hat{a}_{i,k}+\frac{1}{\lambda}\sum_{i=1}^L\hat{\sigma}_i^x\sum_kg_k(\hat{a}_{i,k}+\hat{a}_{i,k}^\dagger).
\end{aligned}
\end{equation}
This Hamiltonian expresses the quantum spin-1/2 chain, each site of which is in contact with its own boson bath.
The Lindblad equation in the singular-driving limit reads
\begin{equation}
\begin{aligned}
\dv{t}\rhoS=-i\left[\Heff-\frac{\eta_+-\eta_-}{2}\sum_{i=1}^L\hat{\sigma}_i^z,\rhoS\right]
+\gamma_+\sum_{i=1}^L\left(\hat{\sigma}_i^+\rhoS\hat{\sigma}_i^--\frac{1}{2}\{\hat{\sigma}_i^-\hat{\sigma}_i^+,\rhoS\}\right)\\
+\gamma_-\sum_{i=1}^L\left(\hat{\sigma}_i^-\rhoS\hat{\sigma}_i^+-\frac{1}{2}\{\hat{\sigma}_i^+\hat{\sigma}_i^-,\rhoS\}\right),
\end{aligned}
\label{eq:Lindblad_many-body_example}
\end{equation}
where the effective Hamiltonian is now given by
\begin{equation}
\Heff=\sum_{i=1}^L\left(J\hat{\sigma}_i^z\hat{\sigma}_{i+1}^z+h^z\hat{\sigma}_i^z+\frac{\xi}{2}\hat{\sigma}_i^x\right),
\label{eq:Heff_many-body}
\end{equation}
and the constants $\gamma_\pm$ and $\eta_\pm$ are respectively given by \cref{eq:gamma_pm,eq:eta_pm}.

\Cref{eq:Lindblad_many-body_example} can have a nontrivial steady state that is not written in the form of Gibbs state $\rho_\mathrm{G}=\e^{-\beta_\mathrm{eff}\Heff}/\TrS[\e^{-\beta_\mathrm{eff}\Heff}]$ for any choice of $\beta_\mathrm{eff}$.
This fact is clearly seen by considering the case of $h^z=0$.
In this case, the system Hamiltonian [\cref{eq:Heff_many-body}] corresponds to the transverse-field Ising model, which possesses the $Z_2$ symmetry $(\hat{\sigma}_i^x,\hat{\sigma}_i^y,\hat{\sigma}_i^z)\to(\hat{\sigma}_i^x,-\hat{\sigma}_i^y,-\hat{\sigma}_i^z)$ for all $i$.
As a result of the $Z_2$ symmetry, the thermal expectation value of $\hat{\sigma}_i^z$ is zero at any finite value of $\beta$: $\TrS[\hat{\sigma}_i^z\rho_\mathrm{G}]=0$.
In contrast, \cref{eq:Heff_many-body} breaks the $Z_2$ symmetry when $\eta_+\neq\eta_-$ or $\beta_+\neq\beta_-$, and hence the steady state $\rho_\mathrm{ss}$ of \cref{eq:Heff_many-body} has a nonzero value of $\TrS[\hat{\sigma}_i^z\rho_\mathrm{ss}]$. 
This fact clearly shows that the steady state of \cref{eq:Heff_many-body} is not given by the thermal Gibbs state.

\section{Discussion}
\label{sec:summary}

In this paper, we have introduced the singular-driving limit, which allows us to derive the Lindblad equation that is valid even in the strong dissipation regime.
Remarkably, the derived Lindblad equation can have a nontrivial steady state in contrast to that in the conventional singular-coupling limit.
Although we only present a simple example in \cref{sec:example}, theoretical framework presented in this paper would be used to design nontrivial steady states of open quantum many-body systems in future works.

Theoretically, the strong-coupling thermodynamics has received recent interests~\citep{Seifert2016, Jarzynski2017}, and the result of this work should be relevant to this general problem.
Without the driving field, the steady state under strong coupling to an equilibrium bath is described by \emph{the mean-force Gibbs state}~\citep{Kirkwood1935, Jarzynski2004, Mori2008,Cresser2021}
\begin{equation}
\rho_\mathrm{MFG}=\frac{\TrB\left[\e^{-\beta(\HS+\HB+\HI)}\right]}{\Tr_\mathrm{SB}\left[\e^{-\beta(\HS+\HB+\HI)}\right]},
\end{equation}
which is nothing but the reduced density matrix of the Gibbs state of the total system.
The mean-force Gibbs state describes the correction to the bare Gibbs state $\e^{-\beta \HS}/\TrS[\e^{-\beta\HS}]$ due to the system-bath coupling, and its property has been intensively investigated in recent years.

In this paper, we consider a system under strong coupling to an equilibrium bath \emph{and} the fast periodic driving, the latter of which is expressed as a time-dependent interaction Hamiltonian in the rotating frame.
It is a future problem to give a general expression of the steady-state density matrix in such a situation by extending the notion of the mean-force Gibbs state to the time-dependent interaction Hamiltonian.

\bmhead{Acknowledgements}
This work was supported by JSPS KAKENHI Grant No. JP21H05185 and by JST, PRESTO Grant No. JPMJPR2259.

\bmhead{Data availability}
No datasets were generated or analysed during the current study.

\section*{Declarations}
\bmhead{Conflict of interest}
The author has no relevant financial or non-financial interests to disclose.

\bibliography{physics}% common bib file

\end{document}